# Interactive Summarizing -

## Automatic Slide Localization Technology as Generative Learning Tool

Lili Yan
Instructional Technology and Learning Sciences
Utah State University
Logan, Utah, USA
e-mail: liliyan@aggiemail.usu.edu

Kai Li
School of Communication and Information Engineering
Shanghai University
Shanghai, P. R. China
e-mail: kailee@t.shu.edu.cn

*Abstract*—**Making a summary is a common learning strategy in lecture learning. It is an effective way for learners to engage in both traditional and video lectures. Video summarization is an effective technology applied to enhance learners' summarizing experience in a video lecture. In this article, we propose to apply cutting-edge automatic slide localization technology to lecture video learning experience. An interactive summarizing model is designed to explain how learners are engaged in the video lecture learning process supported by convolutional neural network and the possibility of related learning analytics.**

## I.  INTRODUCTION

Summarizing is a common practice for learners in a lecture. It is an essential strategy of generative learning [1]. By summarizing contents in a lecture, lecture learners involve actively in the sense-making process. For the increasingly popular video lectures, summarizing is an integral part of the learning process. Affordances of the emerging video technology can be employed to create a smart summarizing experience for video lecture learners.

Compared to traditional lecture experience, video lecture attains the advantage of being more self-regulated. Individual choice of time, place and space of learning is allowed [2]. The flexibility of video lecture has been recognized by and utilized in new pedagogical forms such as "flipped" classrooms, "blended" classroom, and massive open online courses ("MOOCs") as the basic forms of conveying contents.

However, there is a recognized lack of interaction in video lecture learning. The recent advancement of video technologies has opened up possibilities to improve the interactivity of video-based learning experience. These applications contribute to an increasingly recognized "learner agency" in video lecture learning. Video summarizing technology provides assistance to the cognitive interaction between learners and the multimodal learning content in video lectures.

In study, we describe the application of video summarizing technology that 1) facilitates the self-paced video lecture learning, 2) helps to build an interactive summarizing model, and 3) provides possibility to use video summarization as analytical tool for learning behaviors.

## II.  LITERATURE REVIEW

### A.  Interactivity in video lecture

Video lectures are a growing resource of learning within and beyond formal school setting. The popularity of video lectures has driven further research on the effective design and the use of video lectures. Researchers have identified video lecture as an alternative to traditional live lecture. It has better [3] or equal [2] effect on learning outcomes.

One critique of video recording of lectures is the lack of interactivity. Traditional lectures are preferred because of the interpersonal interaction [4]. The form of video lecture can be an example of instructionism reinforced by technology. While the properties of video depend on "the creative shaping of designers, editors, and producers" [5], the learning experience associated with video lecture depends on the degree of engagement it enables.

In fact, video lectures also gives rise to a new model of textual engagement. Instructionism is one stereotype of video lectures. Friesen [6] has pointed out that it is biased to say that lectures are "simple transmission of knowledge" (p. 95). The "hermeneutic speech act" (p. 100) of lectures is reinforced by a range of computing technology in recent years ([7], p. 211). Instructor and students are related in same way of the author-reader interactions. While there are existing researches that make use of computer technology to engage students in the course level or interpersonal level, less has been investigated in terms of how to facilitate the engagement between learners and video lecture content, especially from the student-as-reader perspective.

### B.  Applied computer technology in video lectures

Recent research (2010-present) shows that computer technology is leveraged to facilitate video lecture learning. Digital tools or platform have been used to create learner-centered experience. Discussion forum has been recognized as means to enhance interaction for video lecture [8]. Video blogging or vlogging has been studies as an effective module to engage students in a video lecture [9]. Danmaku, a text-based subtitle commentary has been applied in video lectures with confirmed value of enhancing interaction and course engagement in online video lectures [10]. Apart from supporting communication, video technologies were also used in video lecture analytics. Attention-based video lecture review mechanism [11] has been developed to generate video sections for review determined by learners' attention status. The development of video/audio information retrieval technology contributes to the learner-centered browsing experience. Audi technology has enabled an audio key search experience in video lectures [12]. Video indexing with advanced video framework (ICS videos) and standard optical character recognition (OCR) technology is used by various STEM disciplines. Video lecture components that help with generative learning are still in need.

*C. Summarizing as Learning*

Video summarizing technology contributes to the smart component in video lecture learning and facilitates of lecture summarizing process.

Long seen as skill for text passages, summarizing skills are also used by learners for video lectures. The audience/reader analogy explains the similarity of the summarizing process of video lecture, the new "text". Making summary is generative because the process involves actively making-sense of new information [1]. Mayer [13][14] proposed the SOI model (select-organize-integrate) as a subcomponent of the cognitive theory multimedia learning. The application of video summarizing technology is a supplementary to the generative learning experience.

### III. AUTOMATIC SLIDE LOCALIZATION

In this section, we propose a fully automatic system for extracting the semantic structure from a raw lecture video by using the deep learning technologies [15].

*A. Introduction*

Given a lecture video, the main purpose is to find key frames where slide content changes and generate a summary for the video learners [16].

Since the lecture video is usually recorded by pan-tilt-zoom cameras, camera motion, camera switch, and audience movement inevitably happens, which cause a significant video appearance change. In addition, a slide transition always occurs in a short time along with the progressive content changes, which may result in a subtle appearance change of the video.

Apparently, the appearance change in the video does not necessarily indicate a slide transition. It is hard to tell the real slide transition from the disturbance like camera motion. Therefore, it is difficult to automatically generate the knowledge-related summary for learners.

*B. Design Process*

The slide localization problem is converted to a multi-class classification problem in machine learning. By partitioning the video into lots of small segments, we train a neural network to tell whether each segment contains slide transition or not. In this way, it is easy to produce a keyframe summary by gathering the slide transition segments.

The Convolutional Neural Networks (CNN) is a powerful model for computer vision problems. To take advantage of the temporal evolution among adjacent video frames, especially for slide change detection, we apply 3D CNN [17] to learn spatial-temporal features in videos with the convolution kernels extracted from 2D to 3D. However, as the number of stacked layers increase, 3D CNN costs a large amount of GPU memory and is difficult to train. To solve this problem, we introduce the Residual Network architecture [18] to reduce training time and optimize the performance.

Specifically, different types of lecture video are collected from the website according to the camera recording method. Each lecture video is temporally and spatially down-sampled and for fast processing. N frames together are grouped into a frame volume. Each frame volume may be divided into three classes: (1) unchanged slide, including camera motion and speaker movement; (2) camera switch between slide and the speaker; (3) real slide transition. After manually labeling the groundtruth of each lecture video, we train such as slide transition detector with this new network.ß

The trained network is used for testing any other lecture videos. It will automatically select the slide transition frames and summarize it in a proper manner.

*C. Technical Advancements*

We present a spatial-temporal residual network to better characterize the slide changes. Compared to 2D CNN, 3D CNN applys 3D convolution and 3D pooling operations to model temporal information. However, with the increasing network depth, it is not easy to optimize the network as the deep network leads to saturated or degraded accuracy. Fortunately, residual learning reduces the deep network complexity by identity mapping. By combing the 3D CNN and Residual Network together, we construct a network architecture which contains seven convolutional layers and four fully connected layers. This spatial-temporal framework is able to neglect the disturbance such as camera motion, and reveal the real slide change.

### IV. INTERACTIVE SUMMARIZING MODEL

Video summarization has the merit of providing the better engagement between learners and the content. It contributes to the generative learning experience. It is a "constructive and knowledge-building experience [19]." In this section, we describe the specifics of the interactive summarizing process that are enabled by leveraging automatic slides localization technology (see Fig. 1).

*A. Selection*

Selection is the initial stage of interactive summarizing. Critical slides are automatically selected and present to the learner. Learning occurs at learners' own pace. Slides or related video clips can be reviewed when desired.

Furthermore, in the selection stage, learners usually go through the sensory information (audio, video, text etc) and retain what can be connected to their knowledge construction. With the help of automatic slide localization, learners can choose to process less sensory information by focusing on the automatically selected slides. They could also compare the information they select from the original video and the selected slides. The process leads to an effective selection of information and possibly with less cognitive load.

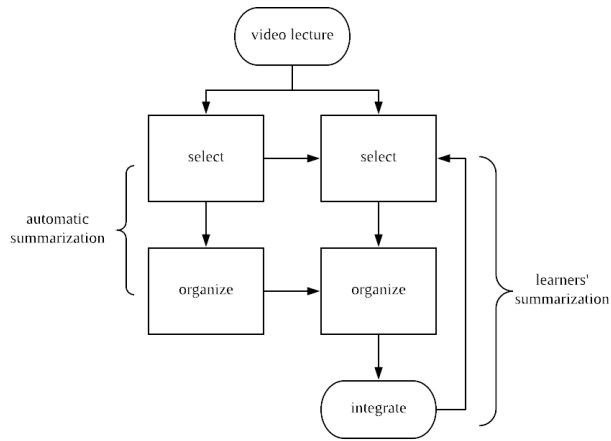

Figure 1. Interactive Summarizing Model.

*B. Organization*

In the organization stage, contents are further organized into a coherent structure. It requires learners' higher level of cognitive ability. By looking at the automatically generated outline, learners' are actively involved in the organizing process. Their evaluation of the automatic organization serves as basis for their generation of the content outline.

*C. Integration*

Integration is stage where a summary of content is produced. A summary usually consists of the main ideas or key messages of the content, expressed in the manner of learners' own words. With the convenience to review information processed in the previous stage, learners' summary production experience is enhanced.

Creating quality summary requires extensive training [1]. The iteration of selecting and organizing process also help to practice and self evaluate summary generation process. In this way, the model also helps to improve learners' summarizing skills, given the situation that the feedback from instructors or peers is not always available.

V. DISCUSSIONS

Interactive summarization model enhances video lecture learning by facilitating learners' summarization process. It therefore opens up the possibility for the analytics of learning process in a video lecture. By examining the electronic behaviors of users, the hidden patterns of learning behaviors could be revealed. It will disclose more about the summarizing process of learners and thus inform the instructor of the critical design of video contents.

ACKNOWLEDGMENT

This work was supported by National Natural Science Foundation of China (No. 61601278), and "ChenGuang" project by Shanghai Municipal Education Commission & Shanghai Education Development Foundation (No. 17CG41).